\documentclass[aps,prl,twocolumn,superscriptaddress]{revtex4-1}
\usepackage{float}
\usepackage{graphicx}
\usepackage{amsmath,amssymb,bbm,mathrsfs,bm,braket,color,graphicx,comment,amsfonts,dsfont}
\usepackage[colorlinks,citecolor=blue,urlcolor=blue]{hyperref}
\usepackage[mathscr]{euscript}
\usepackage[normalem]{ulem}
\usepackage{comment}

\DeclareMathAlphabet\mathbfcal{OMS}{cmsy}{b}{n}

\DeclareMathOperator{\arcsinh}{arcsinh}
\usepackage{extarrows}
\usepackage{mathrsfs}

\begin{document}

\title{Highly Tunable Spin-Orbit Torque and Anisotropic Magnetoresistance in a Topological Insulator Thin Film Attached to Ferromagnetic Layer}

\author{Ali G. Moghaddam}\email{agorbanz@iasbs.ac.ir}
\address{Department of Physics, Institute for Advanced Studies in Basic Sciences (IASBS), Zanjan 45137-66731, Iran}
\address{Research Center for Basic Sciences \& Modern Technologies (RBST), Institute for Advanced Studies in Basic Science (IASBS), Zanjan 45137-66731, Iran}
\address{Institut f{\"u}r Physik, Martin-Luther Universit{\"a}t Halle-Wittenberg, D-06099 Halle, Germany}

\author{Alireza Qaiumzadeh}
\address{Center for Quantum Spintronics, Department of Physics, Norwegian University of Science and Technology, NO-7491 Trondheim, Norway}

\author{Anna Dyrda\l}
\address{Faculty of Physics, Adam Mickiewicz University, ul. Umultowska 85, 61-614 Pozna\'n, Poland}

\author{Jamal Berakdar}
\address{Institut f{\"u}r Physik, Martin-Luther Universit{\"a}t Halle-Wittenberg, D-06099 Halle, Germany}
\date{\today}

\begin{abstract}
We investigate spin-charge conversion phenomena in hybrid structures of topological insulator (TI) thin films and magnetic insulators. We find an anisotropic inverse spin-galvanic effect (ISGE) that yields a highly tunable spin-orbit torque (SOT). Concentrating on the quasiballistic limit, we also predict a giant anisotropic magnetoresistance (AMR) at low dopings. These effects, which have no counterparts in thick TIs, depend on the simultaneous presence of the hybridization between the surface states and the in-plane magnetization. Both the ISGE and AMR exhibit a strong dependence on the magnetization and the Fermi level position and can be utilized for spintronics and SOT-based applications at the nanoscale.
\end{abstract}
\maketitle

\emph{Introduction}.--- The discovery of new types of topological phases and topological insulators (TIs)   \cite{hasan2010review,qi2010review,ryu2016,haldane,witten2016,moore2017,ando} has opened up a new line of fundamental research with prospective applications in electronic and optical devices  \cite{hasan2010review,qi2010review}.
Spin-related phenomena are at the heart of TIs \cite{han2018review,tokura2019review,macdonald2018} due
to the \emph{spin-momentum locking} property of their surface states as gapless excitations protected by time-reversal symmetry (TRS) \cite{hasan2010review,qi2010review,fert2016nature}.
With these properties, TIs can be used to convert pure spin excitation as a carrier of information into an electric (charge) signal or to electrically control magnetization \cite{garate2010,nagaosa2010,nomura2010,tserkovnyak2012,kohno2014,manchon2016prb,manchon2017prb,manchon2019rmp,fert2016,manchon2018,ho2017scirep,ren2018,mou-yang2019,kim2017-hybrid}. 
In most previous studies, only out-of-plane magnetizations effectively influenced the surface states through the generation of a Dirac mass term \cite{manchon2017prb,kohno2014,garate2010,nagaosa2010}. Large spin-orbit torques (SOTs) and the resulting magnetization switching have been demonstrated for hybrid magnetic/TI structures
\cite{mellnik2014,fan2014,wang2015prl,fan2016,kondou2016,han2017prl,mahendra2018,khang2018}.
Other studies have reported the reciprocal effect of spin-electricity signal conversion and spin-pumping with an exceptionally large efficiency \cite{saitoh2014,jamali2015,baker2015,rojas2016}. 

\par
In this letter, we predict a new feature for thin TIs attached to ferromagnetic (FM) layers with in-plane magnetization: When the thickness of the TI approaches a few quintuple layers (QLs), the surface states at the two sides start hybridizing, and a bandgap opens in the surface state spectrum even without a perturbation that breaks the TRS \cite{zhang2010natphys,linder2009,shen2010,liu2010,lu-shen2011,yakovenko2012,balatsky2015,sulaev2015,ghaemi2010}.
Intriguingly, we find that the average of the in-plane magnetizations can also modify the energy dispersion of the surface states. 
This is surprising because for nonhybridized TI surfaces (e.g., for thicker films), the in-plane components of magnetization can be gauged away \cite{burkov2011,parhizgar2016,litvinov2020,bauer17,dyrdal20}, and they do not contribute to effects such as gap opening \cite{chen2010massive,hasan2011disorder,Davis2015pnas,xu2012hedgehog,Ferreira2013}. Then, as a key finding, we illustrate that the interplay of hybridization with the in-plane magnetization significantly influences the SOTs originating from the inverse spin-galvanic effect (ISGE)~\cite{edelstein,manchon2017prb} and the anisotropic magnetoresistance (AMR). For a certain range of chemical potentials feasible for experiments, the current-induced spin densities exhibit large anisotropy, and AMR becomes very prominent. Also, the strong dependence of the spin densities on the magnetization and the chemical potential yield a magnetoelectrically controllable SOT with a nonlinear magnetization dependence, which can be utilized in TI-based spintronic devices, SOT nano-oscillators \cite{chen2016,Akerman}, and even neuromorphic computing, which was recently proposed \cite{torrejon2017neuromorphic}.
\begin{figure}[t!]
    \centering\includegraphics[width=.9 \linewidth]{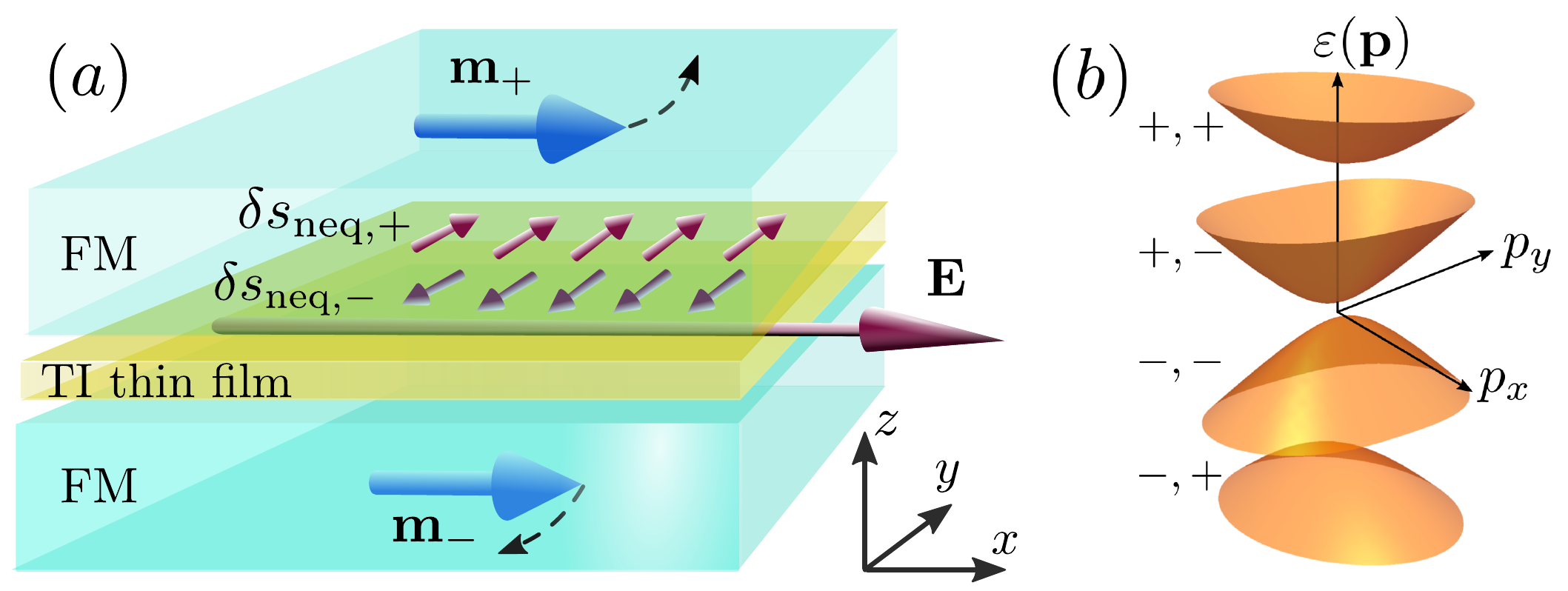}
\caption{(Color online)
The device setup and its band structure. (a) Schematic
of a TI thin film sandwiched between two FM layers. The
ISGE leads to nonequilibrium spin densities perpendicular to the
applied electric field that have opposite directions on the two
surfaces of the thin film.
 (b) Band dispersion
of the thin film assuming $v_F\kappa/\Delta=0.9$. The band labeling is according to the text in the form of a double index $(\nu,\eta).$
}
\label{fig1}
\end{figure}
\par
\emph{Model.}--- We consider a TI thin film with a nanometer-scale thickness, $d$, coupled to one or two adjacent FM layers with magnetizations ${\bf m}_{\pm}$, as schematically shown in Fig.~\ref{fig1}. Assuming Dirac-like surface states at the two sides with hybridization energy $\Delta$, the effective low-energy Hamiltonian of the system is \cite{ghaemi2010,burkov2011,parhizgar2016,litvinov2020,suppmat}
\begin{eqnarray}
\label{H-TI}
{\cal H}=
v_F  \tau_z \otimes (\hat{\bf z}\times {\bm\sigma}) \cdot ({\bf p}-\tau_z {\bm \kappa}-{\bm \kappa}_0)+  \Delta \tau_x \otimes \sigma_0, 
\end{eqnarray}
where $v_F$ denotes the Fermi velocity of the surface states, and the Pauli matrices $\sigma_i$ and $\tau_i$ act in the spin and layer subspaces, respectively. 
Two types of momentum shifts ${\bm \kappa}=(g/2v_F)\sum_\zeta {\bf m}_\zeta \times \hat{\bf z}$ and ${ {\bm \kappa}_0=(g/2v_F)\sum_\zeta \zeta {\bf m}_\zeta \times \hat{\bf z}}$ originate from the exchange coupling ${\cal H}_{\rm ex}=g\sum_\zeta {\bm \sigma}\cdot{\bm m}_\zeta (1+\zeta\tau_z)/2$ between the TI surface states and the magnetization of the adjacent FM layers on top of and beneath the thin film ($\zeta=\pm$ represent the two sides of the surface states). The global momentum shift ${\bm \kappa}_0$ is equivalent to the rigid movement of the full energy bands and can be simply gauged away without any physical consequences. In contrast, the layer-dependent momentum shifts $\pm{\bm \kappa}$ due to the coupling $\Delta$ cannot be gauged out, and as we will see, they radically influence the transport properties of the hybridized surface states \footnote{Detailed explanation of the gauge invariance and symmetry properties of the model can be found in the first section of the Supplemental Material \cite{suppmat}}. 
\par
\emph{Current-induced spin densities.}---
The ISGE or nonequilibrium spin density driven by the charge current originates from spin-orbit coupling.
In TI thin films, the opposite helicities of the surface states at the two sides imply that the current-induced spin densities at the two surfaces also have opposite signs ($\delta {\bf s}_{\rm neq,-}=-\delta {\bf s}_{\rm neq,+}$). Thus, in the linear response, the spin densities
are related to the external electric field ${\bf E}$ as $\delta s_{{\rm neq}, \zeta}^{i} = \zeta (-e) \sum_{j} {\cal S}_{ij} E_j$,
in which ${\mathbfcal S}$ is a second-rank pseudotensor defining the \emph{ISGE response function} (i.e., spin susceptibility). According to the
St{\v{r}}eda-Smr{\v{c}}ka version of the Kubo formula \cite{smrcka1977,streda1982,sinitsyn2006}, the components of ${\mathbfcal S}$ have two contributions related to the Fermi surface and the completely filled energy levels (Fermi sea):
\begin{eqnarray}
\label{1st-ss}
{\cal S}^{\rm I}_{ij}=
\Re 
\int  
\frac{d\varepsilon  d^2{\bf p} }{(2\pi)^3}  
\,  \partial_\varepsilon f(\varepsilon) \,  
&&
{\rm Tr}
\left[ {\hat s}_{i,+} \hat{G}_{\varepsilon}^{R}   {\hat v}_j  ( \hat{G}_{\varepsilon}^{R}- \hat{G}_{\varepsilon}^{A} ) \right],~\\
{\cal S}^{\rm II}_{ij}=
\Re \int  
\frac{d\varepsilon  d^2{\bf p} }{(2\pi)^3}  
~ f(\varepsilon)  ~
&&
{\rm Tr}
\left[ 
{\hat s}_{i,+} 
\hat{G}_{ \varepsilon}^{R}   {\hat v}_j   \partial_\varepsilon\hat{G}_{\varepsilon}^{R}\right.
\nonumber \\
&& 
~~~~~~ \left.
-{\hat s}_{i,+} 
\partial_\varepsilon\hat{G}_{ \varepsilon}^{R}   {\hat v}_j   \hat{G}_{\varepsilon}^{R}) 
\right].~~~~~\label{2nd-ss}
\end{eqnarray}
Here, $f(\varepsilon)$ is the Fermi-Dirac distribution function, and $G^{R,A}_\varepsilon$ denote the momentum-space retarded and advanced Green's functions (the momentum ${\bf p}$ is dropped from $G^{R,A}(\bf{p})$ for brevity). Additionally, $\hat{\bf s}_{\zeta}= (\tau_0+\zeta\tau_z)\otimes{\bm \sigma}/2$ and ${\hat {\bf v}}=v_F\tau_z\otimes{\bm \sigma}$ are the surface-dependent spin operator and velocity operator, respectively. By replacing $s_{i,+}$ in Eqs. (\ref{1st-ss}) and (\ref{2nd-ss}) with $s_{i,-}$, both functions ${\cal S}^{\rm I}_{ij}$ and  ${\cal S}^{\rm II}_{ij}$ also change sign, justifying the appearance of prefactor $\zeta$ in the linear response relation for the spin densities. In addition, the lack of a nontrivial topology and Berry-phase-attributed effects, which is a consequence of the hybridization of the surface states, implies that the contribution from the Fermi sea is negligible \cite{dyrdal2017,dyrdal2019}, and therefore, ${\mathbfcal S}={\mathbfcal S}^{\rm I}$.
\par
The noninteracting Green's function for the clean system defined by Hamiltonian (\ref{H-TI}) is
\begin{eqnarray}
{\hat{G}}_{0\,\omega}^{R,A}({\bf p})=\frac{1}{\omega\pm i0^+ - \hat{\cal H}}  = \sum_{\nu,\eta}\frac{\hat{\mathbfcal P}_{\nu,\eta} ({\bf p}) }
{\omega\pm i0^+-{\varepsilon}_{\nu,\eta} ({\bf p})  },
\end{eqnarray}
in which $\hat{\mathbfcal P}_{\nu,\eta}({\bf p})=
|\psi_{\nu,\eta}({\bf p})\rangle
\langle \psi_{\nu,\eta}({\bf p})|$ is the projection operator to eigenstate $|\psi_{\nu,\eta}({\bf p})\rangle$.
As illustrated in Fig.~\ref{fig1}(b), we have four energy bands ${\varepsilon_{\nu,\eta}({\bf p})=\nu[v_F^2p_x^2 + (v_F\kappa +\eta \sqrt{v_F^2p_y^2+ \Delta^2})^2]^{1/2}}$,
with indices $\nu$ and $\eta$ taking two values $\pm1$.
The projection operators are the sum of the two terms \cite{suppmat}
\begin{eqnarray}
\label{eventerm}
&&\hat{\mathbfcal P}^{({\rm even})}_{\nu,\eta}({\bf p})=\frac{1}{4}[\tau_0\sigma_0
-\nu\cos\theta_\eta \tau_0\sigma_x
\nonumber\\
&&~~~~~~~~~~~ -\eta\,{\rm sech}\,\xi 
\left(
\tau_x\sigma_x
-\nu \cos\theta_\eta \tau_x\sigma_0
\right)],~~\\
&&\hat{\mathbfcal P}^{({\rm odd})}_{\nu,\eta}({\bf p})=\frac{\nu}{4}
[\sin\theta_\eta \tau_z\sigma_y
- \eta\,{\rm sech}\,\xi 
 \sin\theta_\eta \tau_y\sigma_z
\nonumber\\
&&~~-\eta\tanh\xi
\left(\nu
\tau_z\sigma_0
- \cos\theta_\eta \tau_z\sigma_x
+ \sin\theta_\eta \tau_0\sigma_y
\right)],~
\end{eqnarray}
which are even and odd functions of the momentum.
Parameters $\xi=\arcsinh(v_F p_y/\Delta)$ and $\theta_\eta=\arcsin[-v_Fp_x/\varepsilon_{+,\eta} ({\bf p})]$ are used for brevity.
\par
In the presence of disorder, we expect a level broadening matrix $\hat{\bm \Gamma}_\omega$ represented by the imaginary part of the corresponding self-energy function $\hat{\bm \Sigma}_\omega$. Considering short-range impurities with an effective constant potential $V_0$ and a density $n_{\rm imp}$, the level broadening can be expressed within the Born approximation (BA) as
\begin{eqnarray}
\hat{\bm \Gamma}_{\omega}&&=\Im {\bm \Sigma}^{\rm BA}_{\omega}=n_{\rm imp} V_0^2  \int \frac{d^2{\bf p}}{(2\pi\hbar)^2} ~ \Im{\hat{G}}_{0\omega}^{R} ({\bf p}),\nonumber\\
&&=- \frac{\gamma}{\pi}   \int v_F^2  d^2{\bf p}   \sum_{\nu,\eta}  \delta[\omega -\nu{\varepsilon}_{\eta} ({\bf p})] \Re \hat{\mathbfcal P}_{\nu,\eta} ({\bf p}), \label{gamma-p}
\end{eqnarray}
in which the dimensionless impurity scattering strength is given by
$\gamma =n_{\rm imp} V_0^2/(2 \hbar v_F)^2$. By integration over momentum, only even terms ${\mathbfcal P}^{({\rm even})}_{\nu,\eta}({\bf p})$ do not vanish; therefore, one obtains the decomposed form $\hat{\bm \Gamma}_{\omega}=\sum_{i,j=0,1}{\Gamma}_{ij}\tau_i\otimes\sigma_j$, following from Eq. (\ref{eventerm}).
Interestingly, due to their similar matrix structure, the terms $\Gamma_{00}$,
$\Gamma_{01}$, and $\Gamma_{10}$ can be absorbed into
$\omega$, $\kappa$, and $\Delta$ in the expressions for the retarded Green's function:
\begin{eqnarray}
\label{renormalize}
{\hat{G}}_{\omega}^{R}({\bf p})=\left[\omega - \hat{\cal H} -i \hat{\bm\Gamma}_\omega\right]^{-1} =
\left[\widetilde{\omega} - {\widetilde{\cal H}}- i \hat{\bm\Gamma}^\prime_\omega\right]^{-1},
\end{eqnarray}
with the substitutions
${\widetilde{\cal H}} [\kappa , \Delta] \equiv {\cal H} [ \kappa-i\Gamma_{01}/v_F,  \Delta +i \Gamma_{10} ]$,
$\widetilde{\omega}=\omega - i\Gamma_{00}$,
and $\hat{\bm \Gamma}^\prime_{\omega}=\Gamma_{11}\tau_1 \otimes \sigma_1$. The advanced Green's function follows from $\hat{G}^{A}_{\omega}=\hat{G}^{R \,\ast}_{\omega}$.
\par
\begin{figure}[tp!]
    \centering\includegraphics[width=.85\linewidth]{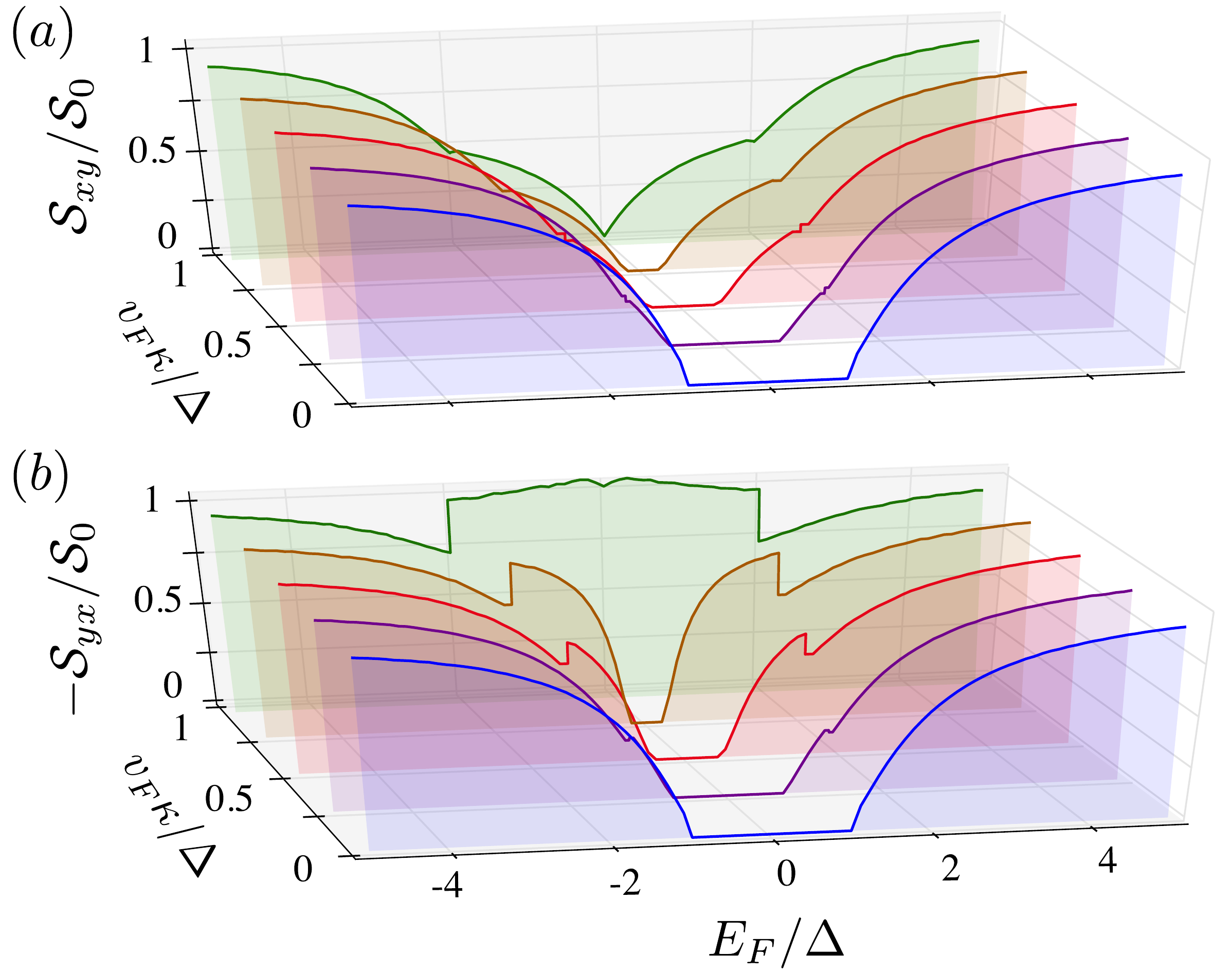}
\caption{(Color online) Energy dependence of the transverse spin-current response functions (a) ${\cal S}_{xy}$ and (b) ${\cal S}_{yx}$ at zero temperature and for different magnetic momentum shifts $\kappa$. Finite values of $\kappa$ give rise to anisotropy in the spin-current response, which becomes profound for $v_F \kappa\sim\Delta$.
Here, we have used $\gamma=0.1$ for the effective disorder strength.
}
\label{fig3}
\end{figure}
\emph{Numerical Results and Discussion}.--- Combining Eq. (\ref{renormalize}) for $\hat{G}^{R,A}_{\omega}({\bf p})$ with Eq. (\ref{1st-ss}),
taking the zero temperature limit with $\partial_\varepsilon f(\varepsilon)=-\delta(\varepsilon)$, and numerically performing the integration, the Fermi level contribution to the spin-current response function, $\mathcal{S}^{I}_{ij}$, is obtained.
We immediately see that only off-diagonal terms ${\cal S}_{xy}$ and ${\cal S}_{yx}$ are nonvanishing, as expected from the chiral form of the low-energy surface state spectrum. Intriguingly, from Fig. \ref{fig3}, the amplitudes of the two components of the spin-current response function can be quite different, indicating the anisotropic nature of the current-induced nonequilibrium spin density. In fact, the anisotropy of the energy bands that originates from the magnetic momentum shift ${\bf\kappa}$ causes a marked difference between ${\cal S}_{xy}$ and ${\cal S}_{yx}$. This difference is particularly evident for the range of Fermi energies $|E_F-\Delta|<v_F|{\bf\kappa}|$, that is, when only one band crosses the Fermi energy. When the Fermi energy falls inside the gap ($E_F< |\Delta- v_F|{\bf\kappa}||$), the induced spin densities identically vanish. In contrast, for large energies, the anisotropy becomes negligible, and the spin-current responses monotonically increase, approaching a constant value ${\cal S}_0=1/ (8\pi v_F \gamma)$ for $E_F\gg \Delta, v_F|\kappa|$ \cite{suppmat}.

\par
\begin{figure}[tp!]
    \centering\includegraphics[width=.86 \linewidth]{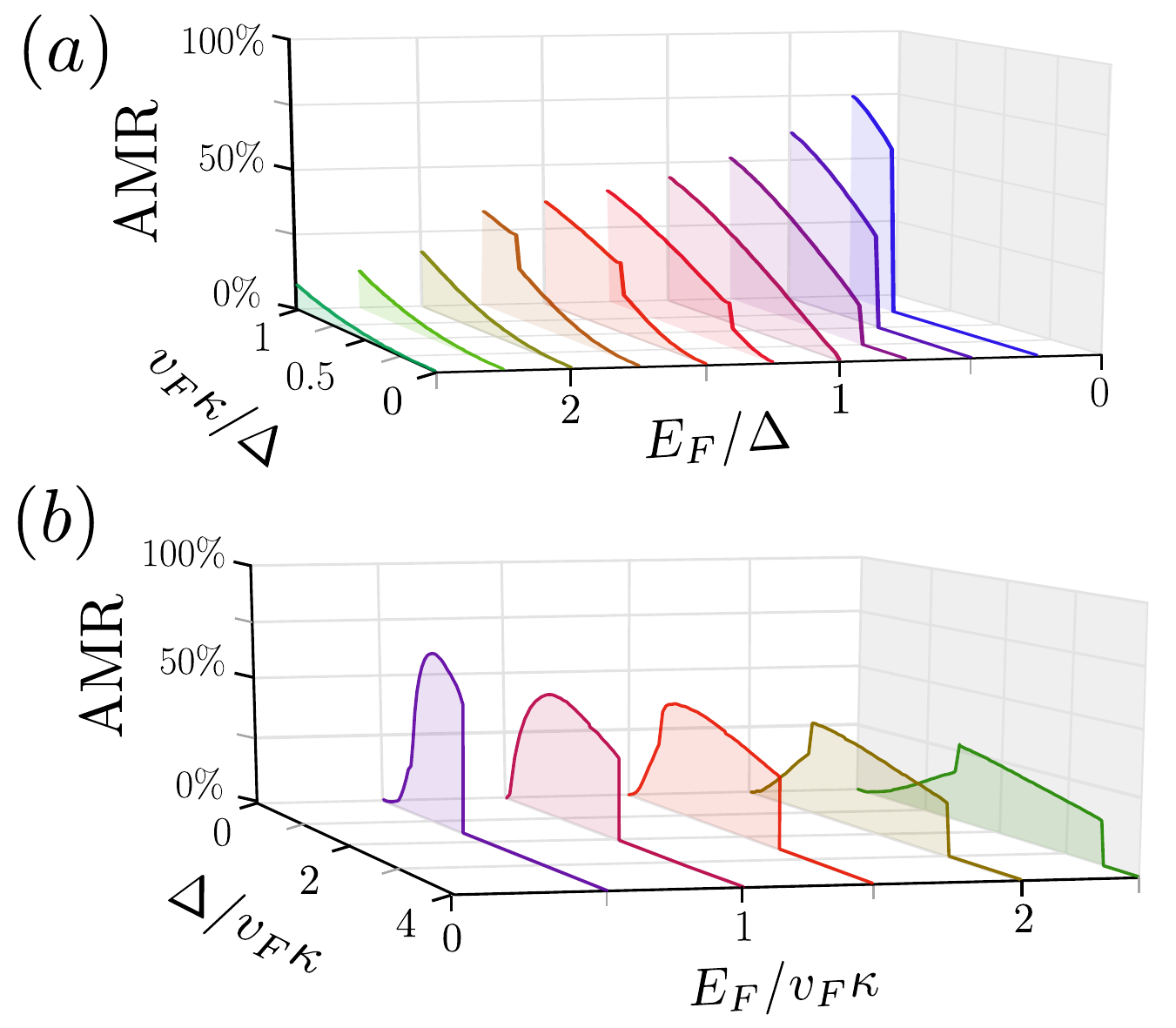}
\caption{(Color online) Variation in the AMR (the percentage of the relative difference between the two components of the longitudinal resistance) with (a) the magnetization of the FM layer and (b) the hybridization of surface states for various values of the Fermi energy.
Note that both horizontal axes in panel (a) have reversed directions for the sake of clarity of the results.}
\label{fig5}
\end{figure}

 
Assuming a weak scattering regime ($\gamma\ll 1$), we find that the spin-current responses decline with the disorder strength as $1/\gamma$, which resembles a conductivity-like behavior. In fact, a relationship between the off-diagonal spin-charge response functions and the diagonal components of the conductivity matrix as ${{\cal S}_{xy}=(\hbar/2e^2v_F)\sigma_{xx}}$ and ${{\cal S}_{yx}=-(\hbar/2e^2v_F)\sigma_{yy}}$, which has been previously found for a single surface of a TI \cite{garate2010,nagaosa2010}, also holds for coupled surfaces. Therefore, anisotropic behavior similar to that shown in Figs.~\ref{fig3} (a) and (b) is expected
for the magnetoconductivities. Figure~\ref{fig5}(a) presents the variation in the AMR with $\kappa$ (note that $\kappa$ defines the magnetization in energy units, $v_{F} \kappa = g m/2$) and the chemical potential. A very large AMR is achieved for small chemical potentials compared to both the hybridization and magnetic energy scales ($E_F\ll\Delta,v_F|\kappa|$). For a fixed chemical potential with respect to $\Delta$, the AMR is generally an ascending function of magnetization. Additionally, the AMR vanishes inside the bandgap and abruptly changes when the number of bands crossing the Fermi level changes. As a result, both the gap width and the splitting between bands of a TI thin film coupled to an FM layer (or in the presence of a magnetic field) can be determined by measuring the AMR. It should be stressed that in Fig. \ref{fig5}, only positive chemical potentials and magnetizations are shown, but due to the symmetry, the same behavior can be expected when changing the sign of each or both of them. Now, as illustrated in Fig. \ref{fig3}(b) for weak overlaps ($\Delta< v_F\kappa$), the AMR is small, and upon approaching $\Delta=0$, it vanishes. This is indeed consistent with the fact that for completely isolated surface states, the opposite momentum shifts $\pm{\bm \kappa}$ can be independently gauged away \cite{suppmat}, and therefore, we do not expect any anisotropy.  

\par
Associated with the nonequilibrium spin densities, the SOT exerted on the two magnetizations on opposite sides can be obtained from ${{\bm\tau}_{\rm SOT,\zeta}=(g/\hbar){\bf m}_\zeta\times \delta {\bf s}_{\rm neq,\zeta}}$.
Then, by choosing the local coordinates such that the in-plane magnetization aligns in the $+\hat{\bf x}$ direction, as shown in Fig. \ref{fig1}, we use the above results for the spin-current response functions and obtain the SOT as follows:
\begin{eqnarray}
{\bm\tau}_{\rm SOT,\zeta}=- \frac{\zeta ge}{\hbar} \:  {\bf m}_\zeta \cdot {\bf E} \: {\cal S}_{yx}   \: \hat{\bf z} . 
\label{SOT}
\end{eqnarray}
Note that ${\cal S}_{yx}$ is a function of the magnetizations due to the momentum shift ${\bm \kappa}$, i.e., ${\cal S}_{yx}=  {\cal S}_{yx}(\sum_{\zeta'}{\bf m}_{\zeta'})$.
This result is in agreement with the general form of the Rashba SOT in two-dimensional systems, especially when only in-plane magnetization is considered \cite{titov2017, ado2020-anisotropic}. Nevertheless, the explicit form of the magnetization dependence is particularly different from those obtained in bulk TI/FM systems when the bandgap is opened by the out-of-plane component of the magnetization \cite{manchon2017prb}.
\par
\begin{figure}
    \centering\includegraphics[width=.85 \linewidth]{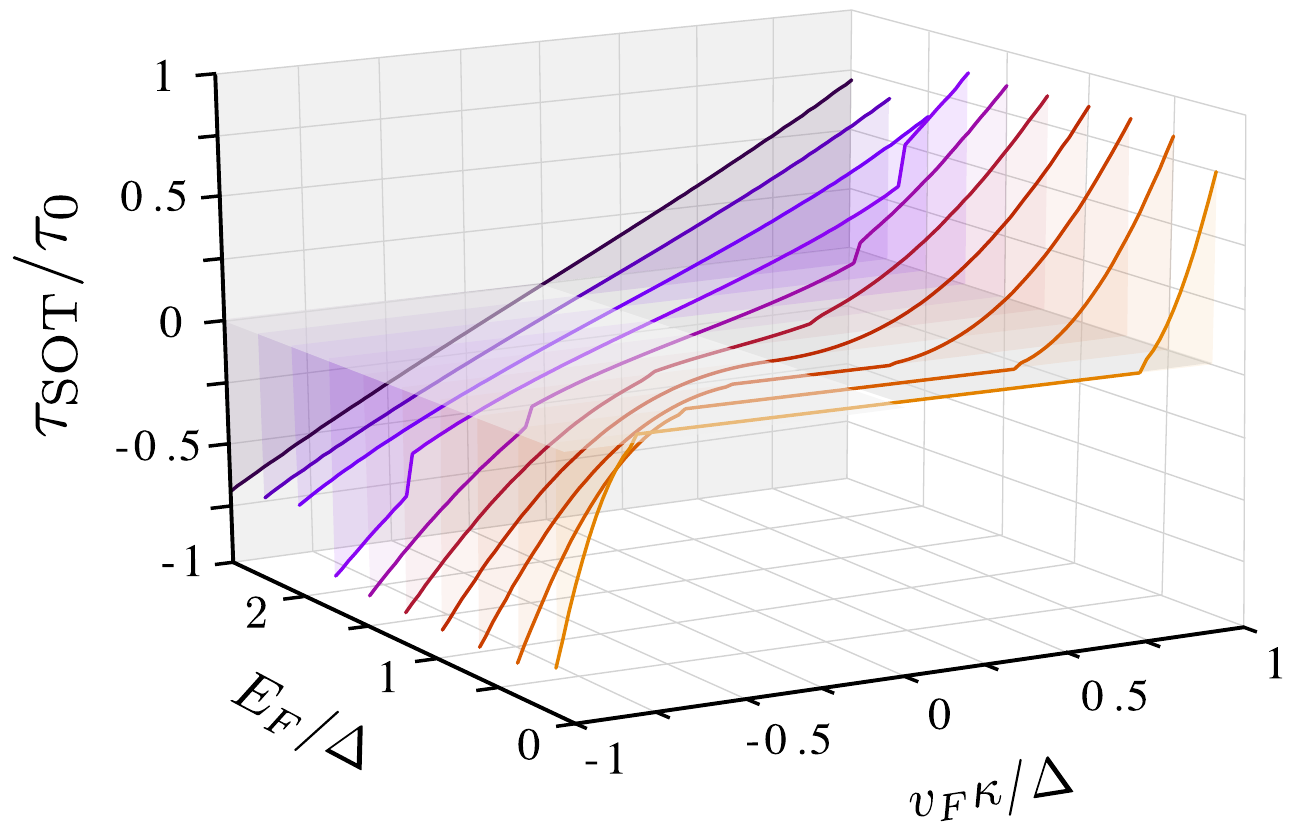}
\caption{(Color online) Variation in the SOT (scaled by $\tau_0=2e E_{\parallel} \Delta {\cal S}_0/\hbar$) with the magnetization of the FM layer for various values of the Fermi energy of the TI thin film. For smaller Fermi energies ($E_F/\Delta \lesssim 1$), the SOT vanishes for the range of magnetic exchange energies $g m/2 \equiv v_F \kappa$, where the Fermi level lies inside a gap, and then starts to increase in a nonlinear manner above a certain value.
At higher energies, the nonlinearity becomes negligible.
}
\label{fig4}
\end{figure}

The variation in the SOT with the magnetization and Fermi energy is shown in Fig. \ref{fig4}. 
Similar to the AMR, the SOT exhibits a significant behavioral change when the Fermi energy moves from the energy gap into the first conduction/valence subbands or reaches the edge of the second conduction/valence bands, suggesting that the bandgap and band splitting can be inferred from the Fermi energy dependence of the SOT.
Importantly, for sufficiently small Fermi energies $E_F\lesssim \Delta$, the SOT nonlinearly varies with the magnetization direction ${\bf m}_\zeta$, which is a direct consequence of the dependence of ${\cal S}_{yx}$ on ${\bm \kappa}$. Therefore, the results of Fig. \ref{fig4} reveal that the SOT in the TI thin film can be magnetoelectrically tailored, which means that one can substantially tune the SOT and magnetic dynamics by changing both the equilibrium magnetization (its amplitude and angle) and the chemical potential via a gate voltage.
This result indicates further advantageous features of FM/TI systems for spintronics and magnetization reversal applications, particularly in comparison with recent elaborate proposals \cite{sinova2017, maghrebi2019}.
\par
The particular advantage of a TI thin film depends on the interplay of the in-plane magnetization and the surface state hybridization, which is manifested in the built-in anisotropy and nonlinearity of the SOT given by Eq. (\ref{SOT}). Various experiments and ab initio calculations have already indicated $\Delta \sim 10-100$ meV for thicknesses of a few QLs \cite{xue2012thin,thinfilm-stroscio-exp,thinfilm-Kim-dft,zhang2010natphys}.
Also, surface gaps due to the out-of-plane magnetization have been reported to vary from a few to $50$ meV in magnetically doped TIs \cite{Davis2015pnas,chen2010massive,hasan2011disorder}. 
Therefore, combined with the Fermi level tunability by doping and external gating, 
our findings should be experimentally realizable, especially for the parameter range $E_F\sim \Delta\sim v_F\kappa$.

\par
Throughout the paper we have focused on in-plane magnetizations, to highlight their role in the thin film limit of FM/TI hybrids. The addition of out-of-plane components does not influence the anisotropy of the bands; instead, this addition augments the tunability of the transport properties by changing the band splittings. 
As a consequence, the qualitative features of ISGE, SOT, and AMR remain intact, although they change quantitatively. Moreover, a damping-like SOT term arises as ${\bm\tau}^{\parallel}_{\rm SOT,\zeta} =-  (\zeta ge/\hbar ) \:  \hat{\bf z} \cdot {\bf m}_\zeta \:  {\cal S}_{yx}   \: {\bf E} $, which has been extensively explored in the case of thick TI/FM \cite{mellnik2014,fan2014,wang2015prl,fan2016,kondou2016,han2017prl,kohno2014,manchon2016prb,manchon2017prb}.
\par
\emph{Conclusions}.--- In this work, we have revealed unprecedented transport features of  
ferromagnetic/topological insulator thin films, namely large anisotropy and a strong dependence on doping in current-induced nonequilibrium spin densities and the resulting spin-orbit torques. 
These effects originate from the role of exchange coupling in the in-plane magnetization in the presence of a finite overlap between the surface states, which has been overlooked until now. 
We further predict a very large AMR in the low doping limit, which diminishes for thick TI structures with
negligible hybridization of the surface states. Given that both the SOT and the AMR depend on the doping and the magnetization, thin films of FM/TI offer promising applications in spintronics and SOT-based devices.

\emph{Acknowledgments}.---
A.G.M. acknowledges financial support from DFG through Grant SFB TRR 227 and thanks MLU Halle-Wittenberg for the hospitality during his visit.
A.Q. has been supported by the European Research Council via Advanced Grant No. 669442 ``Insulatronics'', the Research Council of Norway through its Centres of Excellence funding scheme, Project No. $262633$,  ``QuSpin'', and the Norwegian Financial Mechanism 2014-2021 (Project No. 2019/34/H/ST3/00515, ``2Dtronics'').
A.D. acknowledges the financial support from the National Science Center in Poland (NCN) project No. DEC-2018/31/D/ST3/02351.

\bibliography{mqdb.bib}
\end{document}


\title{Supplemental material}

\title{Supplemental material: \\ ``Highly Tunable Spin-Orbit Torque and Anisotropic Magnetoresistance in a Topological Insulator Thin Film Attached to Ferromagnetic Layer"}

\author{Ali G. Moghaddam}
\address{Department of Physics, Institute for Advanced Studies in Basic Sciences (IASBS), Zanjan 45137-66731, Iran}
\address{Research Center for Basic Sciences \& Modern Technologies (RBST), Institute for Advanced Studies in Basic Science (IASBS), Zanjan 45137-66731, Iran}
\address{Institut f{\"u}r Physik, Martin-Luther Universit{\"a}t Halle-Wittenberg, D-06099 Halle, Germany}

\author{Alireza Qaiumzadeh}
\address{Center for Quantum Spintronics, Department of Physics, Norwegian University of Science and Technology, NO-7491 Trondheim, Norway}

\author{Anna Dyrda\l}
\address{Faculty of Physics, Adam Mickiewicz University, ul. Umultowska 85, 61-614 Pozna\'n, Poland}
\address{Institut f{\"u}r Physik, Martin-Luther Universit{\"a}t Halle-Wittenberg, D-06099 Halle, Germany}

\author{Jamal Berakdar}
\address{Institut f{\"u}r Physik, Martin-Luther Universit{\"a}t Halle-Wittenberg, D-06099 Halle, Germany}

\begin{abstract}
In this Supplemental Material, we provide more details on the analytical results and also general properties of the model.
The first section is devoted to the discussion on the gauge symmetric structure of the model. Then for the general case with finite magnetic exchange $\kappa$, we construct the projection operators which is used to find the zeroth order Green's function as well as level broadening due to disorder. Then for the special case of vanishing magnetic exchange coupling ($\kappa=0$), we present the explicit and analytical derivation of the level broadening function, the Green's function and finally the Edelstein response.
\end{abstract}

\maketitle

\section{Gauge symmetry of the model}

In this part, we rigorously illustrate the origin and differences of two momentum shifts ${\bm\kappa}$ and ${\bm \kappa}_0$ in the Hamiltonian of the hybrid system of a topological insulator (TI) thin film coupled to ferromagnetic (FM) insulators. The low-energy Hamiltonian of an isolated TI thin film can be written as \cite{shen2010,linder2009}
\begin{eqnarray}
{\cal H}_{0}=\begin{pmatrix}
v_F   (\hat{\bf z}\times {\bm\sigma}) \cdot {\bf p} & \Delta \\
\Delta & -v_F(\hat{\bf z}\times {\bm\sigma}) \cdot {\bf p}
\end{pmatrix}. \label{bare-thin-film}
\end{eqnarray}
Here, for the sake of clarity, the matrix structure of the Hamiltonian with respect to the two surfaces is explicitly presented, whereas in the main text we use Pauli matrices $\tau_i$ to indicate the operation in the subspace of the two side layers.
the minus sign for the lower surface states is needed to satisfy the inversion symmetry between the two surfaces, as $\hat{\bf z}\to -\hat{\bf z}$ under inversion symmetry operation.

The exchange coupling between the electrons at the surface of TI thin film and the local moments of FM layers on top of and beneath the thin film reads
\begin{eqnarray}
{\cal H}_{\rm ex}=g\begin{pmatrix}
 {\bm \sigma}\cdot {\bf m}_+ & 0 \\
0 &  {\bm \sigma}\cdot {\bf m}_- \\
\end{pmatrix},
\end{eqnarray}
which is of a short-range $s-d$ origin and each surface is only exchange coupled to its neighboring FM layer.
Using the identity 
${\bm \sigma}\cdot {\bf m} = (\hat{\bf z}\times {\bm\sigma}) \cdot (\hat{\bf z}\times {\bf m}) + m_z\sigma_z$ which is valid for any vector ${\bf m}$, the in-plane components of magnetic exchange coupling can be absorbed into the Dirac Hamiltonian of the two surfaces. So the Hamiltonian ${\cal H}={\cal H}_{0}+{\cal H}_{\rm ex}$ of the hybrid system is obtained as
\begin{eqnarray}
{\cal H}= 
\begin{pmatrix}
 h_{+}({\bf p})+ m_{+,z}\sigma_z & \Delta \\
\Delta & h_{-}({\bf p})+ m_{-,z}\sigma_z
\end{pmatrix}
~,\label{full-H}
\end{eqnarray}
with 
\begin{eqnarray}
h_{\pm}({\bf p}) &&= \pm v_F (\hat{\bf z}\times {\bm\sigma}) \cdot ({\bf p} \pm 
\frac{g}{v_F}  \hat{\bf z}\times {\bf m}_{\pm})\nonumber \\
&&= \pm v_F (\hat{\bf z}\times {\bm\sigma}) \cdot ({\bf p} - {\bm \kappa}_{\pm}).
\label{full-dirac}
\end{eqnarray}
So the Dirac Hamiltonian of two surfaces encounters  
two different momentum shifts ${\bm \kappa}_{\zeta}=-\zeta  \hat{\bf z} \times {\bf m}_{\zeta} = \zeta {\bf m}_{\zeta} \times \hat{\bf z}$ ($\zeta=\pm1$) which can be decomposed in terms of 
\begin{eqnarray}
&& {\bm \kappa} = \frac{{\bm \kappa} _+ - {\bm \kappa} _-}{2}= \frac{g}{2v_F} ({\bf m}_++ {\bf m}_-) \times \hat{\bf z} ,\\
&& {\bm \kappa} _0 = \frac{{\bm \kappa} _+ + {\bm \kappa} _-}{2}= \frac{g}{2v_F} ({\bf m}_+ - {\bf m}_-) \times \hat{\bf z} ,
\end{eqnarray}
using which the Dirac Hamiltonians (\ref{full-dirac}) can be rewritten as  
\begin{eqnarray}
h_{\pm}({\bf p}) = \pm v_F (\hat{\bf z}\times {\bm\sigma}) \cdot ({\bf p} 
-{\bm\kappa_0} \mp {\bm\kappa}).\label{hpm}
\end{eqnarray}
Above expressions readily show that ${\bm \kappa}_0$ is nothing but a global shift of momenta in Hamiltonian and it can be simply gauged out by a transformation 
\begin{eqnarray}
\Psi({\bf r})\to  e^{i {\bm \kappa}_0\cdot {\bf r}} \: \Psi({\bf r})~. \label{gauge}
\end{eqnarray}
In other words, the presence of ${\bm \kappa}_0$ has no effect on the shape of energy bands and eigenstates and subsequently, on the transport properties which are obtained by integration over all momenta. In contrast, the other term ${\bm \kappa}$ corresponds to the opposite shift of the momenta of states at the two surfaces and
provided by the finite hybridization of the surface states, the momenta shifts $\pm{\bm\kappa}$ in (\ref{hpm}) cannot be gauged out. Therefore, these terms which only depend on the sum of the in-plane magnetizations of the FM layers become physically relevant 
and can influence the band structure and transport properties of the hybrid FM/TI thin film. The fact that only sum of in-plane magnetizations ($\sum_\zeta m_\zeta \times \hat{\bf z}$), and not their difference, has physical effect can be also attributed to the inversion symmetry of the Hamiltonian (\ref{bare-thin-film}) in which the opposite side surface states have opposite helicities. 

We have already mentioned that the physical effects attributed to ${\bm\kappa}$ also rely on the presence of hybridization between the surface states. 
To see this, let us consider a thick enough TI films
in which the hybridization between surface states is absent ($\Delta=0$) and therefore states at the top and bottom of TI are completely isolated from each other. In this case, the opposite momentum shifts $\pm{\bm\kappa}$ for the topological surface states in the two sides can be gauged out by performing another transformation
\begin{eqnarray}
\Psi({\bf r})\to  e^{i \tau_z {\bm \kappa}\cdot {\bf r}} \: \Psi({\bf r})~, \label{gauge-2}
\end{eqnarray}
with $\tau_z$ operating in the two layers subspace.
It is clear that the hybridization term in Hamiltonian (\ref{bare-thin-film}) which can be denoted as $\Delta \tau_x$ does not commute with $\tau_z$. Therefore, in the presence of a finite overlap ($\Delta\neq0$), the substitution (\ref{gauge-2}) is no longer a gauge transformation and accordingly, the energy bands shape also changes due to the interplay of overlap $\Delta$ and the exchange term $\kappa$ as discussed in the main text. 
It worth noting that similar features for the gauge invariance have been discussed in the presence of an in-plane external magnetic field, ${\bf B}$, applied to a TI thin film, leading to a momentum shift of the form
${\bm \kappa}_{B}=(1/2) e {\bf B} d \times \hat{\bf z}$ \cite{burkov2011}. The formal similarity between this case and our Hamiltonian can be traced back to the spin-momentum locking of TI surface states.

So far, we have discussed the global gauge symmetry (\ref{gauge}) of the Hamiltonian (\ref{bare-thin-film}) which can be exploited to get rid of global momentum shifts ${\bm \kappa}_0$ as well as an extra gauge transformation (\ref{gauge-2}) which is meaningful only in the absence of $\Delta$ and for isolated surfaces of TI. But it is worth noting that these gauge symmetries are, in fact, limited to the low-energy Hamiltonian which include only linear term in momenta. By inclusion of higher order corrections ${\cal O}({\bf p}^2)$ to the Hamiltonian (\ref{bare-thin-film}), the transformation (\ref{gauge}) generates other terms which depend on ${\bm \kappa}_0$ and cannot be omitted. As a result, at higher energies and momentum ranges when the quadratic and higher order terms in ${\bf p}$ starts to appear as corrections to the low-energy dispersion of the system, the gauge symmetry is broken. Apart from this limitation, at low enough energies when the Dirac-like picture of the surface states is justified and most of the interesting physics show up at this limit, the gauge structure of the Hamiltonian is exact.






\section{Projection operator and level broadening}

\subsection{Derivation of projection operator}
Considering the Hamiltonian,
\begin{eqnarray}
\label{H-TI}
{\cal H}=
v_F  \tau_z \otimes (\hat{\bm z}\times {\bm\sigma}) \cdot ({\bf p}-\tau_z {\bm \kappa})+  \Delta \tau_x \otimes \sigma_0, 
\end{eqnarray}
for the TI thin film at the presence of the magnetic momentum-shift ${\bm \kappa}=\kappa\hat{\bf x}$, we find four sets of eigenvectors $|\psi_{\nu,\eta}({\bf p})\rangle$ given by,
\begin{eqnarray}
&&|\psi_{\nu,\eta}({\bf p})\rangle=\frac{1}{\sqrt{ 2+2 e^{2\eta\xi} }}
\begin{pmatrix}
e^{i\theta_\eta}\\
-\nu\\
\nu \eta e^{\eta\xi}\\
-\eta e^{\eta\xi+i\theta_\eta}
\end{pmatrix},
\end{eqnarray}
which correspond to the four energy bands 
$\varepsilon_{\nu,\eta}=\nu[v_F^2p_x^2 + (v_F\kappa \pm \sqrt{v_F^2p_y^2+ \Delta^2})^2]^{1/2}$.
Here we have $\sinh\xi=v_F p_y/\Delta$, 
$\sin\theta_\eta=-v_Fp_x/\varepsilon_{+,\eta}({\bf p})$ and
each index ($\nu$ and $\eta$) takes two possible values of $\pm 1$. 
Now we can immediately obtain the four projection operator corresponding to each eigenstate, as the following:
\begin{eqnarray}
&& \hat{\mathbfcal P}_{\nu,\eta}=
\frac{-1}{ 2(e^{-\xi} + e^{\xi}) } \nonumber
\\
&& \quad \:\:\: \times 
\begin{pmatrix}
-e^{-\eta\xi } & \nu  e^{i\theta_{\eta}-\eta\xi} & -\nu \eta e^{i \theta_{\eta}} & \eta \\
\nu  e^{-\eta\xi-i \theta_{\eta}} & -e^{-\eta\xi } & \eta & -\nu \eta e^{-i \theta_{\eta}} \\
-\nu \eta e^{-i \theta_{\eta}} & \eta & -e^{\eta \xi } & \nu  e^{\eta \xi -i \theta_{\eta}} \\
 \eta  & -\nu \eta e^{i \theta_{\eta} } & \nu  e^{\eta \xi+i \theta_{\eta} } & -e^{\eta \xi } \end{pmatrix},~~
\end{eqnarray}
We can divide this expression into two parts denoted as $\hat{\mathbfcal P}^{\rm (even)}_{\nu,\eta}({\bf p})$ and $\hat{\mathbfcal P}^{\rm (odd)}_{\nu,\eta}({\bf p})$: the first one includes all the terms which are even with respect to both momenta ($p_x$ and $p_y$), but the second one includes the remainders which are odd at least with respect to either $p_x$ or $p_y$. The significance of this decomposition underlies in the fact that only $\hat{\mathbfcal P}^{\rm (even)}_{\nu,\eta}({\bf p})$ yield nonvanishing result upon integration over ${\bf p}$, especially for calculating the level broadening function ${\bm \Gamma}_{\omega}$. By direct inspection and noting that $\xi$ and $\theta_\eta$ change sign under ${\bf p}\to -{\bf p}$, we obtain the following expressions for the even and odd parts of the projection operators, 
\begin{widetext}
\begin{eqnarray}
&& \hat{\mathbfcal P}^{\rm (even)}_{\nu,\eta}({\bf p})=
\frac{1}{ 4} 
\begin{pmatrix}
1 & -\nu  \cos \theta_{\eta} & \nu \eta \:{\rm sech} \xi \cos \theta_{\eta}  & -\eta \:{\rm sech} \xi \\
-\nu  \cos \theta_{\eta} &1 & -\eta \:{\rm sech} \xi  & \nu \eta \:{\rm sech} \xi \cos \theta_{\eta}  \\
\nu \eta \:{\rm sech} \xi \cos \theta_{\eta} & -\eta \:{\rm sech} \xi  & 1 & -\nu  \cos \theta_{\eta}\\
-\eta \:{\rm sech} \xi  & \nu \eta \:{\rm sech} \xi \cos \theta_{\eta}  
& -\nu  \cos \theta_{\eta} & 1 \end{pmatrix},~~\\
&& \hat{\mathbfcal P}^{\rm (odd)}_{\nu,\eta}({\bf p})=
\frac{\nu}{ 4} 
\begin{pmatrix}
-\nu \eta \tanh\xi & \eta \tanh\xi e^{i\theta_{\eta}}  -i \sin \theta_{\eta} & 
i\eta \:{\rm sech} \xi \sin \theta_{\eta}
 & 0 \\
 \eta \tanh\xi e^{-i\theta_{\eta}}  +i \sin \theta_{\eta}  &-\nu\eta \tanh\xi &
0 & -i \eta \:{\rm sech} \xi \sin \theta_{\eta}  \\
-i \eta\:{\rm sech} \xi \sin \theta_{\eta}
& 0 & \nu\eta \tanh\xi & 
 i \sin \theta_{\eta} -\eta \tanh\xi e^{-i\theta_{\eta}} 
\\
0  & i\eta \:{\rm sech} \xi \sin \theta_{\eta}
& 
 \eta \tanh\xi e^{i\theta_{\eta}}+i \sin \theta_{\eta} 
& \nu \eta \tanh\xi \end{pmatrix}.~~
\end{eqnarray}
\end{widetext}
Using the Pauli matrices $\sigma_i$ and $\tau_i$ we can readily re-write the above expressions in the forms presented in the main text.
%
%
\subsection{Level broadening and spectral function}
\begin{figure}[tp!]
    \centering\includegraphics[width=.95 \linewidth]{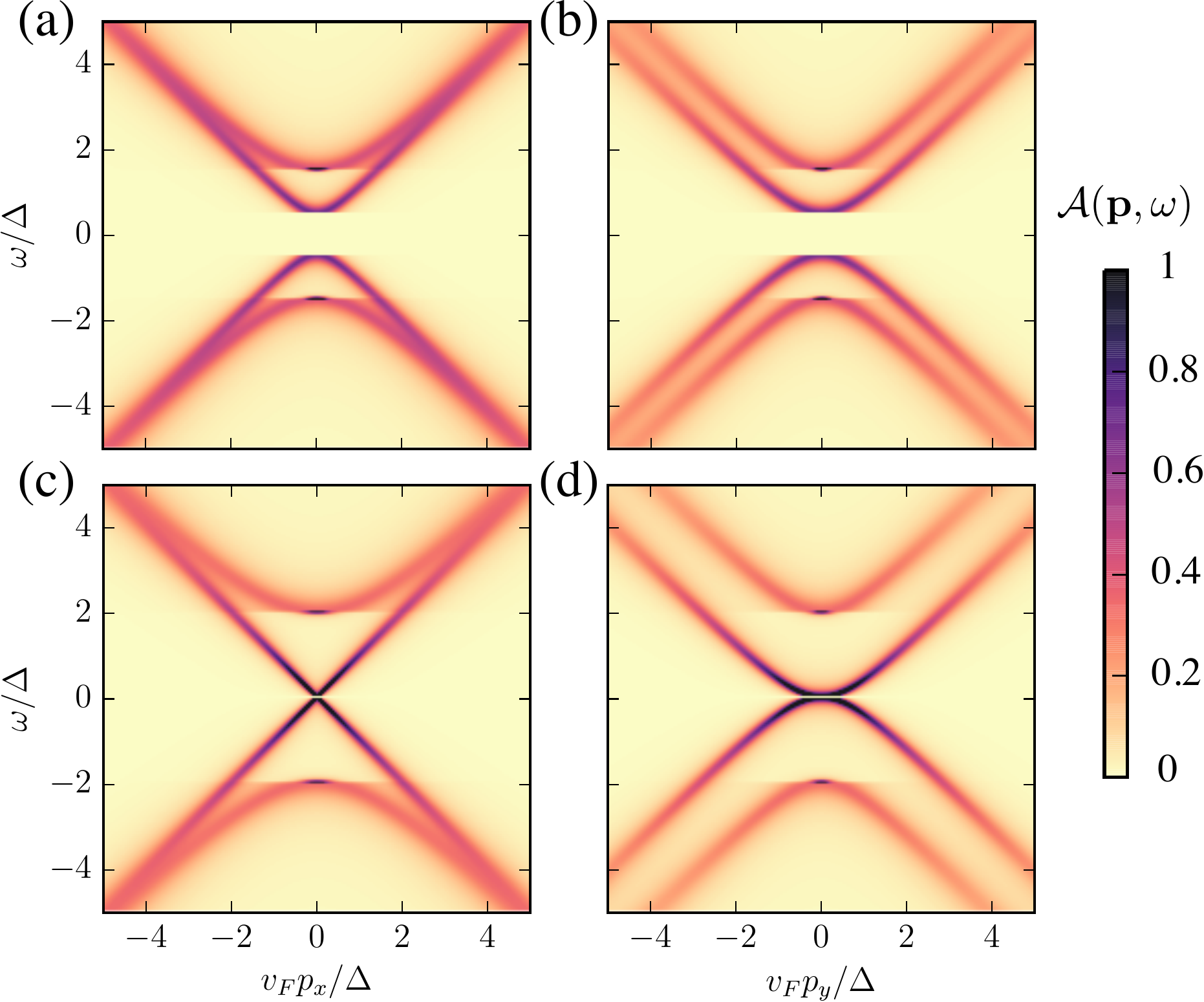}
    \caption{(Color online) Spectral function of the TI thin film coupled to FM layers with in-plane magnetization 
    and at the presence of disorder. (a), (b) show the spectral function for magnetic momentum-shift of $v_F\kappa/\Delta=0.5$ 
    and (c), (d) are corresponding results for $v_F\kappa/\Delta=1$. Left and right panels illustrate dependence on $p_x$ and $p_y$, 
    by setting $p_y=0$ and $p_x=0$, respectively. Here we are using an arbitrary unit scale where the maximum of ${\cal A}({\bf p},\omega)$ 
    is always scaled to unity. The dimensionless impurity scattering parameter which control the level broadening width is $\gamma=0.1$.    
}
\label{fig1supp}
\end{figure}

To illustrate the influence of impurity-induced level broadening besides the band structure of the system, 
Fig. \ref{fig1supp} shows the results of numerical evaluation of the spectral function ${\cal A}({\bf p},\omega)=\Im {\rm Tr}\hat{G}^R_{\omega}({\bf p})$ 
for two different values of magnetic momentum-shift $\kappa$.

\section{Analytical results for ${\kappa}=0$}

\subsection{Level broadening function for ${\kappa}=0$}
In the absence of magnetic momentum shift ${\bm\kappa}$ the energy bands becomes degenerate as 
$\varepsilon_{\eta=\pm}({\bf p})=\varepsilon({\bf p})=\sqrt{v_F^2|{\bf p}|^2+\Delta^2}$ and the relations of angles $\theta_{\pm}$ turn to
$\sin\theta_\eta|_{\kappa=0}=-v_Fp_x/\varepsilon({\bf p})$
 and
$\cos\theta_\eta|_{\kappa=0}=\eta\sqrt{v_F^2p_y^2+\Delta^2}/\varepsilon({\bf p})$. So due to the irrelevance of the index $\eta$ in the energies, 
we can perform a summation over this index inside the projection operator
which has been decomposed as,
\begin{eqnarray}
&&{\mathbfcal P}^{({\rm even})}_{\nu,\eta}({\bf p})=\frac{1}{4}[\tau_0\sigma_0
-\nu\cos\theta_\eta \tau_0\sigma_x
\nonumber\\
&&~~~~~~~~~~~ -\eta\,{\rm sech}\,\xi 
\left(
\tau_x\sigma_x
-\nu \cos\theta_\eta \tau_x\sigma_0
\right)],~~\label{eventerm-old}\\
&&{\mathbfcal P}^{({\rm odd})}_{\nu,\eta}({\bf p})=\frac{1}{4}
[\nu\sin\theta_\eta \tau_z\sigma_y
- \nu\eta\,{\rm sech}\,\xi 
 \sin\theta_\eta \tau_y\sigma_z
\nonumber\\
&&~~-\eta\tanh\xi
\left(
\tau_z\sigma_0
-\nu \cos\theta_\eta \tau_z\sigma_x
+\nu \sin\theta_\eta \tau_0\sigma_y
\right)],~\label{oddterm-old}
\end{eqnarray}
Then summing over $\eta$ drops off the terms which are odd with respect to it and we obtain
\begin{eqnarray}
\sum_{\eta}{\mathbfcal P}^{({\rm even})}_{\nu,\eta}({\bf p})&=&\frac{1}{2}
(
\tau_0\sigma_0
+\nu\eta\,{\rm sech}\,\xi 
 \cos\theta_\eta \tau_x\sigma_0
)\nonumber\\
&=&
\frac{1}{2}
(
\tau_0
+\frac{\Delta}{\nu\varepsilon({\bf p})}
 \tau_x)
 \sigma_0
~~\label{eventerm}\\
\sum_{\eta}{\mathbfcal P}^{({\rm odd})}_{\nu,\eta}({\bf p})&=&\frac{\nu}{2}
(
\sin\theta_\eta \tau_z\sigma_y
+\eta\tanh\xi\cos\theta_\eta \tau_z\sigma_x
)\nonumber\\
&=&
\frac{v_F}{2}
\tau_z \frac{p_y  \sigma_x
-p_x  \sigma_y}
{\nu\varepsilon({\bf p})},
~~
\label{oddterm}
\end{eqnarray}
By plugging this result in the general relation of level broadening function, it can be obtained as what follows:
\begin{eqnarray}
{\bm\Gamma}_{\omega}&&=- \frac{\gamma}{\pi}   \int v_F^2  d^2{\bf p}   \sum_{\nu,\eta}  \delta[\omega -\nu{\varepsilon}_{\eta} ({\bf p})] \Re \hat{\mathbfcal P}_{\nu,\eta} ({\bf p}),\nonumber\\
&&=- \frac{\gamma}{2\pi}   \int v_F^2  d^2{\bf p}  
 \sum_{\nu}  \delta[\omega -\nu{\varepsilon}({\bf p})]
\left(\tau_0
+\frac{\Delta \tau_x}{\nu\varepsilon({\bf p})  } \right) \sigma_0
 \nonumber\\ 
&&=-  \gamma    \int_0^{\infty}  \varepsilon d\varepsilon  
 \sum_{\nu}  \delta(\omega -\nu{\varepsilon}) \left(\tau_0\sigma_0
+\frac{\Delta \tau_x\sigma_0}{\nu\varepsilon } \right)\nonumber\\ 
&&=-\gamma |\omega| \left(\tau_0\sigma_0 +\frac{\Delta}{\omega} \tau_x\sigma_0\right).
\label{gamma-final}
\end{eqnarray}
We note that in passing from the second to the third line of the expression, we have
used the polar representation as $d^2{\bf p}=pdpd\phi_p$ and then implemented a variable change $p\to\varepsilon=\sqrt{v_F^2 p^2+\Delta^2}$. The expression (\ref{gamma-final}), despite the difference in the matrix structure, is reminiscent of the impurity self-energy for an isolated TI surface coupled to an FM system with nonvanishing out-of-plane magnetization $m_z$, which gives rise to gap opening \cite{garate2010,manchon2017prb}.

\subsection{Green's functions for ${\kappa}=0$}
The bare Green's function of TI thin film in the absence of magnetic terms (${\bm \kappa}=0$) can be deduced using Eqs. (\ref{eventerm}) and (\ref{oddterm}) for the projection operator, yielding  
\begin{eqnarray}
&&{\hat{G}}_{0\,\omega}^{R}({\bf p})=\frac{1}{\omega\pm i0^+ - \hat{\cal H}} = \sum_{\nu}\frac{\sum_\eta\hat{\mathbfcal P}_{\nu,\eta} ({\bf p}) }
{\omega\pm i0^+-\nu{\varepsilon} ({\bf p})  }\nonumber\\
&&
=
\sum_{\nu}\frac{
[
\nu\varepsilon({\bf p}) \tau_0
+\Delta \tau_x
]
\sigma_0
+
v_F\tau_z (p_y  \sigma_x
-p_x  \sigma_y)}
{2\nu{\varepsilon} ({\bf p}) [\omega + i0^+-\nu{\varepsilon} ({\bf p}) ]}.
\end{eqnarray}
From the last result we define the following bare Green's function 
\begin{eqnarray}
G_{0,\omega}= \frac{\omega +  v_F  \tau_z (\hat{\bm z}\times {\bm\sigma}) \cdot {\bf p}+  \Delta \tau_x  \sigma_0}{\omega^2-v_F^2|{\bf p}|^2-\Delta^2}, 
\end{eqnarray}
from which the retarded and advanced Green's functions can be obtained using
\begin{eqnarray}
G_{\omega}^{R,A}=G_{0,\omega}[\omega\to\widetilde{\omega}^{R,A},\Delta\to\widetilde{\Delta}^{R,A}], 
\end{eqnarray}
and substitution of $\omega$ and $\Delta$ with $\widetilde{\omega}^{R,A}=\omega[1\mp i\gamma{\rm sgn}(\omega) ]$ and 
$\widetilde{\Delta}=\Delta[1 \pm i\gamma{\rm sgn}(\omega) ]$ respectively, to accommodate the self-energy terms (\ref{gamma-final}).
Assuming $\omega>0$ the impurity-averaged Green's functions read,
\begin{eqnarray}
&&G_{\omega}^{R,A}=\nonumber\\ 
&&~~~\frac{\omega(1\mp i\gamma) +  v_F  \tau_z (\hat{\bm z}\times {\bm\sigma}) \cdot {\bf p}+  \Delta (1\pm i\gamma) \tau_x  \sigma_0}
{\omega^2(1\mp i\gamma)^2-v_F^2|{\bf p}|^2-\Delta^2(1\pm i\gamma)^2},\label{greens} 
\end{eqnarray}
For $\omega<0$ above results can be still used by changing the sign of $\gamma$ as well.

\subsection{The spin-current response for ${\kappa}=0$}

Now after some algebra using the expressions (\ref{greens}),
the integrand of the Kubo formula for the Fermi level contribution in the Edelstein response function is found as, 
\begin{eqnarray}
&& \Xi_{xy,\zeta} \equiv
{\rm Tr} 
\left[ {\hat s}_{x,\zeta} \hat{G}_{\varepsilon}^{R}   {\hat v}_y  ( \hat{G}_{\varepsilon}^{R}- \hat{G}_{\varepsilon}^{A} ) \right]~\nonumber\\
&&~~~
=4\zeta v_F \gamma 
\frac{\left(  v_F^2|{\bf p}|^2 -{\cal Q}^\ast \right) {\cal K}     -4 i v_F^2 p_y^2 \left(\Delta ^2+\varepsilon ^2\right) }{\left( v_F^2|{\bf p}|^2 -{\cal Q}^\ast\right)^2 \left(v_F^2|{\bf p}|^2 -{\cal Q} \right)}
~\nonumber\\
&&~~~
=4\zeta v_F \gamma \left[
\frac{\cal K}{\left( v_F^2|{\bf p}|^2 -{\cal Q}^\ast\right) \left(v_F^2|{\bf p}|^2 -{\cal Q} \right)} \right.
~
~\nonumber\\
&&~~~~~~~~~~~~~~ -\left.
\frac{ 4 i v_F^2 p_y^2 \left(\Delta ^2+\varepsilon ^2\right) \left(v_F^2|{\bf p}|^2 -{\cal Q} \right) }{\left( v_F^2|{\bf p}|^2 -{\cal Q}^\ast\right)^2 \left(v_F^2|{\bf p}|^2 -{\cal Q} \right)^2} \right],
~
\end{eqnarray}
with parameters ${\cal K} =(\gamma +i) \Delta ^2-(\gamma -i)  \varepsilon ^2 $
and ${\cal Q} =(\gamma +i)^2 \Delta ^2-(\gamma -i)^2  \varepsilon ^2 $.
Then the real part of the above expression which we need, reads
\begin{eqnarray}
\Re\, \Xi_{xy,\zeta}
&&=4\zeta v_F \gamma \left[
 \frac{\Re\,{\cal K}}{\left( v_F^2|{\bf p}|^2 -{\cal Q}^\ast\right) \left(v_F^2|{\bf p}|^2 -{\cal Q} \right)} \right.
~
~\nonumber\\
&&~~~~~~~~~ -\left.
\frac{ 4 v_F^2 p_y^2 \left(\Delta ^2+\varepsilon ^2\right) 
 \Im \,{\cal Q}
 }{\left( v_F^2|{\bf p}|^2 -{\cal Q}^\ast\right)^2 \left(v_F^2|{\bf p}|^2 -{\cal Q} \right)^2} \right]
~
~\nonumber\\
&&=4\zeta v_F \gamma^2 \left[
 \frac{ (\Delta ^2-\varepsilon ^2) }{\left( v_F^2|{\bf p}|^2 -{\cal Q}^\ast\right) \left(v_F^2|{\bf p}|^2 -{\cal Q} \right)} \right.
~
~\nonumber\\
&&~~~~~~~~~ -\left.
\frac{ 8 v_F^2 p_y^2 \left(\Delta ^2+\varepsilon ^2\right)^2
  }{\left( v_F^2|{\bf p}|^2 -{\cal Q}^\ast\right)^2 \left(v_F^2|{\bf p}|^2 -{\cal Q} \right)^2} \right].
~~~
\end{eqnarray}
Before we proceed we note that above expression is even with respect to $\gamma$
and therefore, it can be safely used for the case of negative energies without any caution since the aforementioned sign change of $\gamma$ to obtain the result for negative $\varepsilon$ has no effect whatsoever.
Now assuming zero temperature, the Edelstein response function can be written as
\begin{eqnarray}
{\cal S}^{\rm I}_{xy,\zeta}|_{\varepsilon}&=&
-
\int  
\frac{  d^2{\bf p} }{(2\pi)^3} \, 
\Re \, \Xi_{xy,\zeta}
~\nonumber\\
&=&
4\zeta  \gamma^2
\int  
\frac{ { E} d{ E} d\varphi }{(2\pi)^3 v_F} 
[\frac{ (\varepsilon ^2 -\Delta ^2) }
{\left({ E}^2 -{\cal Q}^\ast\right) \left({ E}^2  -{\cal Q} \right)}~\nonumber\\
&&~~~~~~+
\frac{8 { E}^2 \cos^2\varphi (\Delta ^2+\varepsilon ^2 )^2 }
{\left( { E}^2 -{\cal Q}^\ast\right)^2 \left({ E}^2 -{\cal Q} \right)^2}]~,
\end{eqnarray}
where in the second line the integration variable is decomposed in polar coordinates and $E=v_F |p|$. Then we can decompose the two terms inside the integrand versus their basic rational expressions as 
\begin{eqnarray}
&&\frac{1}{{\left( { E}^2 -{\cal Q}^\ast\right) \left({ E}^2 -{\cal Q} \right)}}
~\nonumber\\
&&~~~~~~~~~=\frac{1}{({\cal Q}-{\cal Q}^\ast)}
\left(\frac{1}{E^2-{\cal Q}}-\frac{1}{E^2-{\cal Q}^\ast}\right),
~
\end{eqnarray}
and
\begin{eqnarray}
&&\frac{{ E}^2}{{\left( { E}^2 -{\cal Q}^\ast\right)^2 \left({ E}^2 -{\cal Q} \right)^2}}
~\nonumber\\
&&~~~~~~~~~~~~=\frac{1}{({\cal Q}-{\cal Q}^\ast)^2}
\left[
\frac{{\cal Q}}{\left(E^2-{\cal Q}\right)^2}+\frac{{\cal Q}^\ast}{\left(E^2-{\cal Q}^\ast\right)^2}
\right]
~\nonumber\\
&&~~~~~~~~~~~~-\frac{({\cal Q}+{\cal Q}^\ast)}{({\cal Q}-{\cal Q}^\ast)^3}
\left(\frac{1}{E^2-{\cal Q}}-\frac{1}{E^2-{\cal Q}^\ast}\right),
~
\end{eqnarray}
whose integral over $E$ can be easily performed. Consequently, we obtain the following result
\begin{eqnarray}
{\cal S}^{\rm I}_{xy,\zeta}|_{\varepsilon}
&=&\frac{4\zeta  \gamma^2}{(2\pi)^2 v_F}
\left\{ \frac{1}{4\gamma^2} 
+
\frac{\varepsilon ^2-\Delta ^2}{\varepsilon ^2+\Delta ^2}
\right.
\nonumber\\
&\times &
\left. \frac{1+\gamma^2}{16 i \gamma^3}
 \left[-\log(-{\cal Q})+\log(-{\cal Q}^\ast)\right]
 \right\}
\end{eqnarray}
By inserting $\varepsilon=E_F$, the final result for the spin-current response in the absence of $\kappa$ at upper surface ($\zeta=+1$) becomes 
\begin{eqnarray}
&&{\cal S}_{xy}=-{\cal S}_{yx}= \frac{1}{  (2\pi)^2 v_F  }\left[   
 1-
 (\frac{1}{\gamma_{\rm }}+\gamma_{\rm }){\cal F}(\frac{E_F}{\Delta})
\right],~~\\
&&{\cal F}(x)=\frac{1}{2}\frac{x^2 -1}{x^2 +1} {\rm Arg}\left[ (\gamma-i)^2 x^2 -(\gamma+i)^2 \right],
\end{eqnarray}
with ${\rm Arg}(z)$ indicating the argument of the complex number $z$. Correspondingly, opposite values as above relations hold for the spin-current response in the other surface ($\zeta=-1$). In the quasiballistic limit with very small $\gamma$ and assuming $|E_F|>\Delta$, this result reduces to ${\cal S}_{xy} \approx  {\cal S}_0 (E_F^2-\Delta^2)/(E_F^2+\Delta^2)$,
in which ${\cal S}_0=1/ 8\pi v_F \gamma$.

\bibliography{mqdb.bib}